# A comparative study of the MACD-base trading strategies: evidence from the US stock market

Pat Tong Chio
June 18, 2022

## Abstract

In recent years, more and more investors use technical analysis methods in their own trading. Evaluating the effectiveness of technical analysis has become more feasible due to increasing computing capability and blooming public data, which indie investors can perform stock analysis and backtest their own trading strategy conveniently. Many technical indicators are employed in trading simulations to evaluate the effectiveness of various trading strategies. The Moving Average Convergence Divergence (MACD) indicator is one of the popular technical indicators that are widely used in different strategies. In order to verify the MACD effectiveness, in this thesis, I use the MACD indicator with traditional parameters (12, 26, 9) to build various trading strategies. Then, I apply these strategies to stocks listed on three indices in the US stock market (i.e., Dow-Jones, Nasdaq, and S&P 500) and evaluate its performance in terms of win rate, profitability, Sharpe ratio, number of trades and maximum drawdown. The backtesting is programmed using Python, covering the period between 01/01/2015 and 28-08-2021. The result shows that the win-rate of the strategy with only the MACD indicator is less than 50%. However, the win-rate is improved for the trading strategies that combine the MACD indicator with other momentum indicators like the Money Flow Index (MFI) and the Relative Strength Index (RSI). Based on this result, I redesign the MACD mathematical formula by taking the trading volume and daily price volatility into consideration to derive a new indicator called VPVMA. The results show that the win-rate and risk-adjust performance of this new trading strategy have been improved significantly. Furthermore, I use the Genetic Algorithm (GA) to identify the optimized parameter for the MACD, but it generates mixed results among the stocks listed on different indices. The result implies that there is no single optimized parameter for the MACD indicator. In general, the findings suggest that while all the MACD trading strategies mentioned above can generate positive returns, the performance is not good without using other momentum indicators. Hence, the VPVMA indicator performs better.

# Table of Contents

# List of Figures

# List of Tables

# 1. Introduction

## 1.1. Research Motivation

Trading strategy is an important topic in finance. In general, there are two schools of thought on this topic, including fundamental analysis and technical analysis. Fundamental analysis aims to identify the intrinsic value of a company from its financial statements and justify whether the current stock price is higher or lower than the intrinsic value. Scholars suggest that investors buy the stock when its current stock price is lower than the intrinsic value and vice versa. However, it is difficult to identify the current intrinsic value because financial statements of a company are published quarterly. Therefore, the financial data cannot reflect the company's current situation because of lagging. In addition, the window dressing problem in financial statements is also misleading investors too.

In contrast, technical analysis predicts the price movement based on the trading information such as stock price, trading volume, market news and economic indicators. Investors do not trade stocks based on the financial statement information but rely on price trends. Trends represent that the stock price was going up or going down in a certain period. Investors should buy the stock when it reveals an uptrend and sell the stock in downtrend. Due to the blooming real time data, investors are more confident to execute their customized trading strategy with different technical indicators in the stock market. There is a need for verifying the effectiveness of technical indicators. Hence, this thesis attempts to focus on this important issue.

## 1.2. Research Objective

The MACD is one of the most popular technical indicators. Many indie investors rely on it to build various trading strategies. However, there are many different signal identification rules for the MACD. It indicates that the usage of the MACD is vague, and investors do not know which one is better. Therefore, this thesis aims to verify the effectiveness of MACD strategies, which include different trading rules for MACD, combinations of momentum indicators, MACD parameters optimized and derivation of the MACD indicator.

## 1.3. Research Design

All data used in this thesis are programmatically downloaded from yahoo finance website (https://finance.yahoo.com/)[1] to my backtesting system through its application programming interface (API). For each stock data, yahoo finance provides the following information. Date denotes the date of the stock price; Open denotes the opening market price on that date; High denotes the highest market price on that date; Low denotes the lowest market price on that date; Close denotes the close price on that date; Adj Close denotes the adjustment close price on that date; and Volume denotes the number of trading shares on that date. In this thesis, I use the information mentioned above except for the adjustment closing price, because it reflects the closing price after accounting for any corporate actions. However, it is a theoretical price as no one can buy the stock at Adj Close price in the stock market.

The backtesting system is programmed in Python. I select stocks listed on three indices in the US stock market (i.e., Dow Jones, Nasdaq and S&P 500). The intuitive performance metric is win rate. It tells us the probability of making profit under the selected signal identification rule. However, a successful strategy should not only concern the win rate but also evaluates other aspects. Thus, besides the win rate, P&L ratio, number of trades, Sharpe ratio and Max drawdown should be evaluated too. Specifically, win rate evaluates the probability of making profit, P&L ratio evaluates the proportion of average profit and average amount for the strategy, and the number of trades gives us information about excessive trading. Considering excessive trading increases the transaction cost, the profit will be eroded even if the strategy possesses high win rate and high P&L properties. In addition, Sharpe ratio tells us the risk-adjusted performance and maximum drawdown tells us the maximum loss that investors suffer when the stock price goes from peak to bottom. Hence, the strategy should be evaluated based on the above metrics. The assumptions and criteria of the backtesting process are shown below.

1. Assumptions
    - The initial money for each stock is $80,000

---

[1] https://github.com/ranaroussi/yfinance

- Trade executes after the day of trading signal occurs.
- For a buying signal, the program uses all the money to buy as many as possible shares of the stock. For a selling signal, the program sells all the shares of the current position.
- The calculation only involves the stock data covered in the testing period.
- The minimum number of shares for buying or selling is 1 unit. It complies with the US stock market rule.
- The program does not calculate the transaction cost per trade.
- Shorting behavior is not considered in this thesis.

2. Criteria
    - To consider the win-rate is significant, it should be higher than 50%.
    - Higher P&L ratio represents a high profitability in each trade. Therefore, it should be higher than 1.
    - The strategy should not induce an excessive or insufficient number of trades.
    - Risk and return should be moderate. A high return with higher risk is not considered as an excellent trading strategy. Therefore, a strategy with higher Sharpe ratio should be selected.

Following the above assumptions and criteria, I implement the backtesting system in Python to evaluate the performance of MACD-based strategies. In section 3, I test the MACD strategy with traditional parameters (12, 26, 9) and different signal identification rules firstly. Secondly, I use the GA algorithm to optimize the parameter of MACD and examine its performance. Finally, I combine the MACD with other technical indicators to build 4 strategies and examine their performance. In section 4, I derive a new indicator from the MACD which is VPVMA, and I examine its performance through backtesting. In section 5, I summarize those strategies' results and conclude which strategies are profitable and which strategy is the best among MACD-base strategies.

.

# 2. Technical Indicators

## 2.1 Moving Average Convergence – Divergence (MACD)

The MACD is a trend momentum indicator and was created by Gerald Appel in the late 1970s. It consists of MACD line, which is the difference between fast EMA line and slow EMA line, the signal line, which is the EMA line of MACD and Histogram, which is the difference between MACD line and Signal Line. In general, the parameter of MACD can be adjusted case by case. 12, 26, 9 days combination is the most popular usage in realistic trade. Before you calculate the MACD indicator, you should understand the EMA formula. The mathematical formula is defined as follow:

$$EMA_t = \alpha * closing\ price_t + (1 - \alpha) * EMA_{t-1} \qquad (2.1\text{-}1)$$

where t represents the day of EMA. $\alpha$ represents the degree of decrease; $\alpha = \frac{2}{(t+1)}$.

Higher value decreases faster.

For instance, suppose 12 is the short period and 26 is the long period parameter. The EMA value of 12 and 26 can be calculated by below formulas.

$$EMA_{12} = \frac{2}{(12+1)} * closing\ price_{12} + \left(1 - \frac{2}{(12+1)}\right) * EMA_{12-1}$$

$$EMA_{26} = \frac{2}{(26+1)} * closing\ price_{12} + \left(1 - \frac{2}{(26+1)}\right) * EMA_{26-1}$$

It is the crucial mathematical formula of the MACD. You cannot calculate the result of the MACD without it because the MACD value is the difference between the fast EMA and the slow EMA. The explanation of MACD with (12, 26, 9) parameters in mathematical format is shown as follow:

a) MACD Line: The difference between the fast EMA and the slow EMA.

$$MACD = EMA_{12} - EMA_{26} \qquad (2.1\text{-}2)$$

b) Signal Line: It is the EMA result of MACD Line:

$$Signal = EMA_9 \qquad (2.1\text{-}3)$$

c) Histogram: A bar to represent the difference between MACD line and Signal line.

$$Hist = MACD - Signal \qquad (2.1\text{-}4)$$

When the MACD line crosses above its signal line, the Hist value is greater than 0, otherwise it is below 0. In this thesis, the default parameter of MACD is (12, 26, 9). Therefore, I use the MACD to represent the MACD indicator with (12, 26, 9) parameter over this thesis.

There are many ways to determine the entry time and exit time of the stock trading by applying MACD. I selected four common signal identification rules for evaluating the MACD strategy's performance. They are signal line crossover, zero crossover, histogram and signal line crossover above zero.

a) Signal line crossover: When the MACD line crosses above its signal line, a buy signal is generated. Otherwise, the MACD line crosses below the signal line, a sell signal is generated.

   In mathematical format:

   Buy: $Macd_{t-1} < Signal_{t-1}\ \&\ Macd_t > Signal_t$

   Sell: $Macd_{t-1} > Signal_{t-1}\ \&\ Macd_t < Signal_t$

b) Zero crossover: When the MACD line crosses above zero, a buy signal is generated. As the opposite, when the MACD line crosses below zero, a sell signal is generated.

   In mathematical format:

   Buy: $Macd_{t-1} < 0\ \&\ Macd_t > 0$

   Sell: $Macd_{t-1} > 0\ \&\ Macd_t < 0$

c) Histogram: A buy signal is generated when the most recent three consecutive days are all below zero, but the middle day is the lowest point within three days. In contrast, a sell signal is generated when the most recent three consecutive days are all above zero, but the middle day is the highest point within three days.

   Buy: $H_{t-1} = Minimum\ (H_{t-2}, H_{t-1}, H_t), (H_{t-2}, H_{t-1}, H_t) < 0$

   Sell: $H_{t-1} = Maximum\ (H_{t-2}, H_{t-1}, H_t), (H_{t-2}, H_{t-1}, H_t) > 0$

d) Signal line crossover above zero: When the MACD line crosses above its signal line and MACD line crosses above zero simultaneously, a buy signal is generated. As the opposite, a sell signal is generated.

In mathematical format:

Buy: $Macd_{t-1} < Signal_{t-1}$ & $(Macd_t > Signal_t$ & $Macd_t > 0)$

Sell: $Macd_{t-1} > Signal_{t-1}$ & $(Macd_t < Signal_t$ & $Macd_t < 0)$

## 2.2 Bollinger Bands (BB)

Bollinger Bands was created by John Bollinger in the late 1980s. Bollinger Band consists of three bands. They are lower bands, middle bands and upper bands. The middle band is the 14 days simple moving average (SMA) of the closing price. And the lower band is the difference between SMA and standard deviation. Upper band is the summation of SMA and standard deviation. Usually, investors assume that the standard deviation is 2. The followings are formulas for three bands:

Middle Band = SMA$_{(14)}$ (2.2-1)

Lower Band = Middle Band – Std (2.2-2)

Upper Band = Middle Band + Std (2.2-3)

$$SMA = \frac{1}{14}\sum_{t=1}^{14} P_t \quad (2.2\text{-}4)$$

The trading signal is determined by the SMA of Bollinger Band Width (BBW). It is the normalization value of the difference between upper band and lower band.

$$BBW = \frac{Upper\ Band - Lower\ Band}{Middle\ Band} \quad (2.2\text{-}5)$$

BBW: A buy signal is generated when the SMA (BBW,10) crosses above the SMA (BBW, 50). A sell signal is generated when the SMA (BBW,10) crosses below the SMA(BBW,50).

In mathematical format:

Buy: SMA(BBW,10) > SMA(BBW,50)

Sell: SMA(BBW,10) < SMA(BBW,50)

## 2.3 Relative Strength Index (RSI)

The relative Strength Index (RSI) was created by J. Welles Wilder. It is an oscillator, which measures the price change in a certain period for evaluating an overbought or oversold status of the stock. The whole idea is that investors should buy the stock when it is oversold and sell the stock when it is overbought. Generally, the RSI value is calculated by the past 14 days. The RSI value below 30 indicates the stock is oversold and above 70 indicates the stock is overbought. Usually, investors consider a reversal will occur soon when the RSI value touches the upper value or lower value. The following is the mathematical steps of the RSI:

$$RSI = 100 - \frac{100}{1 + RS} \quad (2.3\text{-}1)$$

$$RS_t = \frac{Average\ Gain_t}{Average\ Loss_t} \quad (2.3\text{-}2)$$

As for the average gain or loss, the calculation step of the first average gain and loss is different to the subsequent average gain or loss.

$$Average\ Gain_0 = \frac{\sum_1^{14} Gain_i}{14} \quad (2.3\text{-}3)$$

$$Average\ Loss_0 = \frac{\sum_1^{14} Loss_i}{14} \quad (2.3\text{-}4)$$

$$Average\ Gain_t = \frac{Average\ Gain_{t-1} * 13 + Gain}{14} \quad (2.3\text{-}5)$$

$$Average\ Loss_t = \frac{Average\ Loss_{t-1} * 13 + Loss}{14} \quad (2.3\text{-}6)$$

The above formula explains that the first average gain or loss is the average of the asset price in the first 14 days, but the afterward average gain or loss is calculated by the previous average and the current average.

Typically, the RSI below 30 represents oversold, and RSI over 70 represents overbought. A buy signal is generated when the stock is oversold, and a sell signal is generated when the stock is overbought. However, these two thresholds limit the investor's profit when the stock price is in a strong uptrend. It indicates that the stock price will be going up even if the RSI is over 70 for a certain period. Therefore, this indicator should be used with other indicators together for capturing the whole rising trend.

Buy: RSI <= 30

Sell: RSI >=70

## 2.4 Money Flow Index (MFI)

The money Flow Index (MFI) was created by Gene Quong and Avrum Soudack. It is an oscillator, which uses stock price and transaction volume to identify the buy signal and sell signal. Same with the RSI indicator, it defines a threshold to determine the stock is overbought or oversold. The whole concept is that investors should buy the stock when it is oversold and should sell the stock when it is overbought. In general, The MFI value below 25 represents oversold and MFI over 70 represents overbought. Compared with the RSI indicator, the MFI indicator takes volume into account. Therefore, people describe MFI as the weight volume RSI.

The followings are the mathematical steps of the MFI:

$$MFI = 100 - \frac{100}{1 + Money\ Flow\ Raito} \tag{2.4-1}$$

$$Money\ Flow\ Ratio = \frac{X\ Period\ Positive\ Money\ Flow}{X\ Period\ Negative\ Money\ Flow} \tag{2.4-2}$$

$$Money\ Flow = Tyipical\ Price * Volume \tag{2.4-3}$$

$$Typical\ Price = \frac{High + Low + Close}{3} \tag{2.4-4}$$

The above depicts that the money flow is considered as positive when the typical price is greater than yesterday's typical price. As the opposite, it is negative money flow.

Typically, the MFI value below 25 represents the stock is oversold and a buy signal is generated. As the opposite, the MFI value over 70 represents the stock is overbought and a sell signal is generated. However, when the stock is in a strong uptrend, the stock price will be going up even if the MFI is over 70. Therefore, making the investment decision based on the MFI indicator will lead to profit limitation too. Hence, this indicator should be used with other indicators together for capturing the whole rising trend.

Trading signal rule:

Buy: MFI <= 25

Sell: MFI >=75

## 2.5 Parabolic Stop and Reverse (SAR)

The SAR indicator was created by J. Welles Wilder Jr. investors use it to trail the price as trend and determine the direction and potential time of trend reverse. It uses a series of dots in the chart to represent the price trend. When the dots are below the stock price, it is trending upward. Conversely, it is trending downward. A potential trade occurs at flip point. A flip point indicates the current trend will be reversed in future. When the stock price is increasing, the SAR value follows the given acceleration value increased. When the stock price is decreasing, the SAR value follows the given acceleration value to decrease too. The mathematical steps of the SAR are shown as follows.

$$SAR_t = SAR_{t-1} + \alpha * (EP - SAR_{t-1}) \text{ uptrend} \tag{2.5.1}$$

$$SAR_t = SAR_{t-1} - \alpha * (SAR_{t-1} - EP) \text{ downtrend} \tag{2.5-2}$$

EP (Extreme point) is the extreme value in the period. When the stock is in uptrend, the EP indicates the highest value in the observed period. As the opposite, The EP is the lowest value in the observed period when the stock is in downtrend.

$\alpha$ is the acceleration factor. Normally, investors use 0.02 as an initial value. $\alpha$ will be updated when the EP has changed. Therefore, when the stock price reaches a higher value during the uptrend period, $\alpha$ should increase too. To prevent $\alpha$ increases too high, a maximum value of $\alpha$ must define. In general, this maximum value is 0.2. But this value can be adjusted by investors case by case. A lower value causes the SAR to be less sensitive to the price volatile, but a higher value causes the SAR to be more sensitive.

Trading signal rule:

Buy: $Sar_{t-1} > Price_{t-1}\ \&\ Sar_t < Price_t$

Sell: $Sar_{t-1} < Price_{t-1}\ \&\ Sar_t > Price_t$

A flipping point of dot is considered a potential trade opportunity. When the dot is above the price but flip below to the price on the next day. A buy signal is generated. Conversely, when a dot flips above the price from below the price on the next day, a sell signal is generated. The flipping sensitivity is influenced by the maximum value of $\alpha$.

# 3. Backtesting

In this section, I show you the backtesting result of different MACD-based strategies. The backtesting process is implemented in Python. All the market data is downloaded from yahoo finance website and the technical indicator value is calculated by TA-Lib library[2]. It is a famous and popular technical indicator library in Python. In the first part of the backtesting result, I measure the performance of MACD strategies with different signal identification rules. In the second part, I use the GA algorithm to identify the optimized parameter for the MACD in the US stock market. Finally, I combine the MACD with aforementioned technical indicators to measure the performance of MACD-based strategies.

## 3.1. Performance Metrics:

The following metrics are used to evaluate the performance of MACD-based strategies. The risk-free rate is assumed 0 because the covering testing period is in the lowest-interest rate environment.

Number of Trades (NT): The total number of trades are executed in the backtesting.

Win ratio (WR): It indicates the probability of profitability on each trade.

$$Win\ ratio = \frac{Profit\ transactions}{Total\ transaction} \qquad (3.1\text{-}1)$$

---

[2] https://mrjbq7.github.io/ta-lib/

Profit and Loss (P&L): It measures the profitability. The higher ratio is better.

$$P\&L = \frac{Total\ Gain}{number\ of\ winning\ trades} \div \frac{Total\ Loss}{number\ of\ loss\ trades} \quad (3.1\text{-}2)$$

Sharpe Ratio (SR): It evaluates the risk adjusted return per unit of risk taken in the strategy.

$$Sharpe\ Ratio = \frac{return\ of\ portfolio - risk\ free\ rate}{standard\ deviation\ of\ excess\ return\ of\ the\ portfolio} \quad (3.1\text{-}3)$$

Sortino Ratio: It evaluates the risk adjusted return over the volatility of downside.

$$Sortino\ Ratio = \frac{return\ of\ portfolio - risk\ free\ rate}{standard\ deviation\ of\ downside} \quad (3.1\text{-}4)$$

Maximum drawdown (MDD): It indicates the decline from peak to trough before a new peak is achieved. It points out the size of losses investors may suffer.

$$Maximum\ Drawdown = \frac{Trough\ Value}{Peak\ Value} - 1 \quad (3.1\text{-}5)$$

## 3.2. Technical Indicators & Trading Strategies

I evaluate 8 different MACD-based strategies and the MACD strategy that adopts optimized parameter. The optimized parameter is generated by the GA algorithm. Four strategies are MACD strategies with different signal identification rules. The rest of them are strategies, which consist of MACD and aforementioned indicators. The result of MACD strategies and the MACD strategy that adopts optimized parameter show that they are profitable, but the risk return is not good. The combination strategies perform better than MACD strategies from the risk return and win rate aspects, but the accumulated profit is not lucrative. All strategies are evaluated by three US stock market indices and their constituents (Dow Jones, Nasdaq and S&P 500) data sets. As a starting point, I show the comparison result of the MACD strategy with different signal rules firstly.

### 3.2.1 Comparison result for MACD strategy with different signal rules

Table 1 reports that the performance of MACD strategies with different signal rules on different US stocks market indices data sets under the traditional MACD parameter (12,26,9).

Table 1: The comparison result of the MACD strategy with different trading signal rules by adopting the traditional parameter.

|  | NT | Win rate | P&L | SR | Sortino ratio | MDD |
|---|---|---|---|---|---|---|
| Panel A: Dow Jones | | | | | | |
| $MACD_{crossoversig}$ | 1939 | 0.41 | 1.97 | 0.33 | 0.51 | 0.28 |
| $MACD_{crossoverzero}$ | 837 | 0.37 | 2.53 | 0.28 | 0.41 | 0.31 |
| $MACD_{hist}$ | 1439 | 0.68 | 0.70 | 0.31 | 0.46 | 0.38 |
| $MACD_{crossoversigabout0}$ | 217 | 0.50 | 2.92 | 0.34 | 0.50 | 0.36 |
| Panel B: Nasdaq | | | | | | |
| $MACD_{crossoversig}$ | 6485 | 0.41 | 2.15 | 0.44 | 0.67 | 0.35 |
| $MACD_{crossoverzero}$ | 2694 | 0.37 | 3.95 | 0.49 | 0.74 | 0.36 |
| $MACD_{hist}$ | 4761 | 0.68 | 0.82 | 0.46 | 0.69 | 0.39 |
| $MACD_{crossoversigabout0}$ | 697 | 0.52 | 4.74 | 0.50 | 0.75 | 0.43 |
| Panel C: S&P 500 | | | | | | |
| $MACD_{crossoversig}$ | 32894 | 0.40 | 2.15 | 0.38 | 0.58 | 0.34 |
| $MACD_{crossoverzero}$ | 13984 | 0.37 | 3.08 | 0.33 | 0.50 | 0.37 |
| $MACD_{hist}$ | 23870 | 0.66 | 0.71 | 0.31 | 0.45 | 0.45 |
| $MACD_{crossoversigabout0}$ | 3616 | 0.49 | 3.23 | 0.35 | 0.52 | 0.43 |

The comparison result shows the MACD strategy with histogram trading rule ($MACD_{hist}$) has the highest win rate among the above strategies for each data set. The number of trades is 1439, 4761 and 23870 respectively but the P&L is less than 1 for each data set. It indicates that the average loss amount is larger than average profit even though the number of trades is larger compared to others. Figure 1 illustrates that the loss amount generated by adopting this strategy is larger than the MACD strategy with signal line crossover rule ($MACD_{crossoversig}$) and the MACD strategy with zero crossover rule ($MACD_{crossoverzero}$), especially in S&P 500. The largest loss amount is more than 80 percent of the principal. As for the gain, the largest is nearly 60 percent. The results correspond to the P&L ratio. This is a small profit strategy.

As for the $MACD_{crossoversig}$ strategy, this strategy generates 1939, 2694 and 32894 transactions in the backtesting for each data set respectively. The number of trades is larger. For instance, 323 trades are executed on the data set of Dow Jones on yearly average. Therefore, this strategy has an excessing trading problem. The MACD$_{crossoverzero}$ strategy performs worse than others in terms of win rate. The win rate of this strategy is 0.37 respectively for data sets, and the Sharpe ratio is 0.28 for Dow Jones data set. It is extremely low. But the number of trades generated by this strategy is lower than the signal crossover rule and histogram rule for all data sets. For instance, 837 number of trades are generated by the MACD$_{crossoverzero}$ for the Dow Jones data sets. Compared to MACD$_{hist}$ and MACD$_{crossoversig}$. It reduces over half of the transactions. The possible explanation for this effect is because the MACD line crosses the signal line frequently. Finally, the MACD strategy with signal line crossover trading rule (MACD$_{crossoversigabout0}$) performs better. The win rate for each dataset is closed to 0.5. Although the number of trades is few, the P&L ratio is the highest among those strategies. For each data set, the P&L ratio is higher than 2. It indicates that profitability is outstanding. However, the Sharpe ratio for constituents of Dow Jones and constituents of S&P 500 are unattractive. They are 0.34 and 0.35 respectively. In addition, the maximum drawdown for each data set is slightly higher. It indicates investors suffer from stock volatility and the proportion between the return and risk is imbalance. All in all, the win rate of the MACD strategy for data sets are mostly below 0.5, and the Sharpe ratios are low. It implies the return that is generated by the MACD strategy is very risky. Is risky return a feature of the MACD? Or is the result of misusing parameters? For further study, identifying the optimized parameter for the MACD applied to stocks listed on US stocks market indices becomes critical.

Figure 1: The scatter plots for the MACD$_{hist}$ strategy.

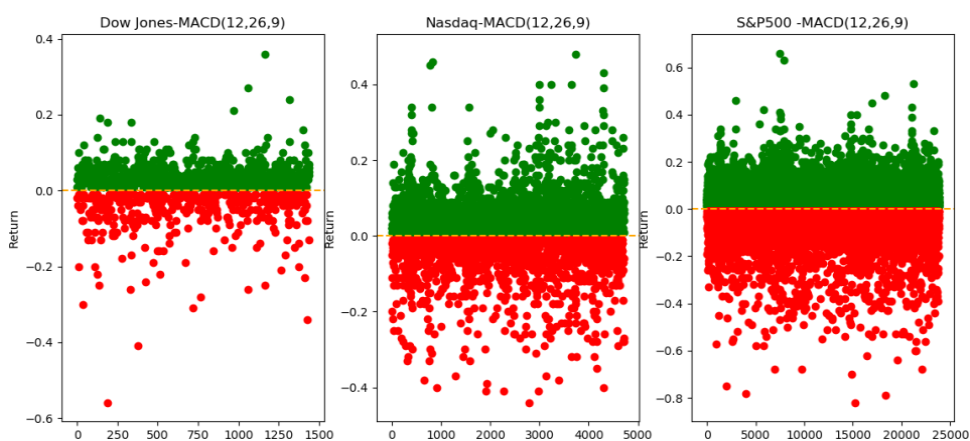

### 3.2.2 Optimization MACD Parameter

The exhaustion search can enumerate all possible candidates for the optimized parameter of MACD. It guarantees that researchers can get the best solution from the given condition. However, the process of enumerating is time consuming under the search range for the following MACD parameter.

(4, 20) for the fast period.

(6, 21) for the slow period.

(4, 41) for the signal period.

The method of exhaustion search is not a good approach to identify the optimized parameter for the MACD under the given range. Instead of exhaustion search, the genetic algorithm (GA) is an algorithm that can identify the optimized parameter for the MACD in a simple and fast approach. The GA is created by John Holland. The process of GA is based on the evolution and genetics principle in Biology. The GA consists of 6 calculation steps in one iteration and repeat the iteration until the fitness value converges, or the iteration reaches the predefined maximum. You can consider the iteration is equivalent to the generation in Biology. The result of each iteration is passed to the next iteration by a container, which is population. Population consists of certain amounts of chromosomes. A Chromosome consists of genes. Gene is the minimum unit in the GA algorithm. During the iteration, chromosomes will experience selection, crossover and mutation. Once the fitness value converges, the GA will stop even if the iteration does not reach the preset number. The fitness value is calculated based on the predefined objective function formula. In general, the best fitness value always indicates the highest value of objective function. As for the selection process, I use the roulette wheel selection approach in order to pass the best gene to the next generation. For simplicity, I use a cutoff approach to implement the crossover. In addition, I use a random drawing approach to implement the mutation. The followings are the definition of the chromosome, the gene, the fitness function, the selection process, the crossover process and the mutation process.

- Iteration: 100. It means the GA will be executed 100 times maximally.
- A gene: It is the basic unit in the GA. A gene means a parameter for the MACD, it represents the fast line parameter, the slow line parameter and the signal parameter.

- Chromosomes consist of genes. It is a tuple type value in Python.
- Population consists of chromosomes and evolves in iteration. The size of population is 30 for this GA.

Objective function: It evaluates the performance of the chromosome. The best chromosome should generate a higher objective value. In this case, the objective function is to maximize the product of profit and the win rate.

$$\max \text{ fitness value} = \text{profit} * \text{win rate} \tag{3.2-1}$$

Selection: This process selects two chromosomes as a pair from the population for producing the offspring. The offspring experiences mutation and passes to the population of the next generation. In this GA, I use the roulette wheel selection approach to select chromosomes.

Crossover: The two selected chromosomes produce the offspring according to the following steps.

Mutation: The offspring is slightly different from their parents. The process for producing the difference is called mutation in the GA. The mutation process is implemented as follows:

Figure 2: The overall illustration for elements of the Genetic Algorithm.

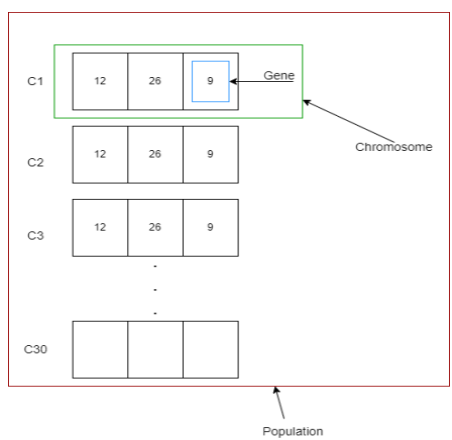

Figure 3: The illustration for crossover process.

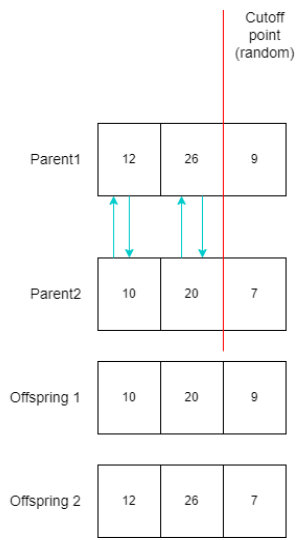

Figure 4: The illustration for mutation process.

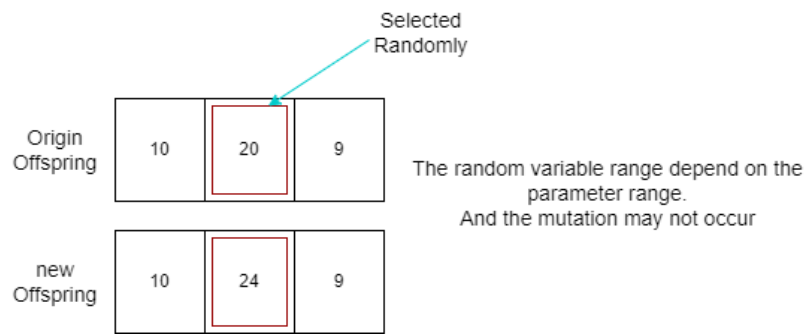

Figure 5: The illustration for the GA process.

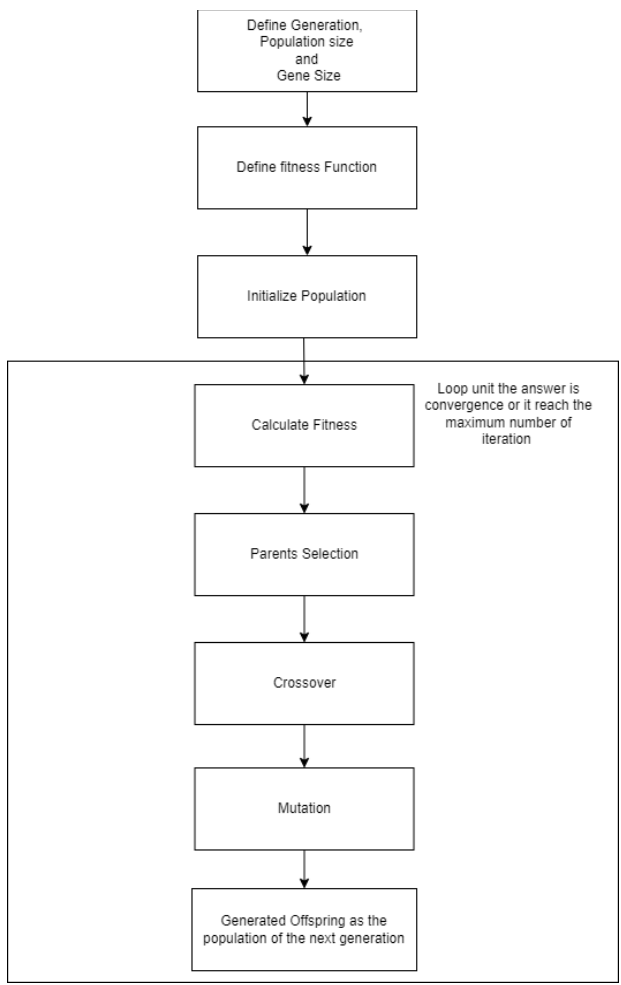

In this part, I use ETFs to run the GA instead of all constituents of indices because it can speed up the process time. The selected ETFs are DIA, QQQ and SPY. They represent Dow Jones, Nasdaq and S&P 500 indices respectively. After the GA generates optimized parameters for the MACD on different ETFs, I apply them to the $MACD_{crossover sig about 0}$ strategy for evaluating the performance.

Table 2: The optimized result for the US stock market indices

| Data set | Parameters |
|---|---|
| Dow Jones | (9,27,40) |
| Nasdaq | (18,35,39) |
| S&P 500 | (17,31,23) |

Table 2 reports the results of the optimized parameter for the US stock market indices respectively. The optimized parameters are different for each data set. This result indicates that there is no single optimal solution for the US stock market. The longer signal parameter is the only attribute that they are in common.

Table 3 reports the comparison results between traditional parameter and optimized parameter. Although the win rate of optimized parameter for the Dow Jones datasets is not improved, Nasdaq and S&P 500 are improved as expected. From the trading performance aspect, the P&L ratio is improved as expected by adopting optimized parameter. A higher P&L value indicates that the average return of each trade outperforms the traditional parameter. The return is lucrative. However, Figure 6 illustrates that the lucrative returns are actually driven by some outliers. These outliers boost the P&L ratio to a high value. Two of the outliers are over twenty times of the initial investment. In addition to this, it is important to note that the US stock market has experienced a bullish market in the testing period. Therefore, a longer position generating a high return is a common phenomenon in the bullish market. Furthermore, the highest P&L ratio is caused by some of the extreme high returns. Once I exclude these return outliers from the observation, the average return is almost equivalent to the traditional one.

The Sharpe ratio of Nasdaq and S&P 500 has been improved. But the improvement for Nasdaq is not outstanding. In addition, the maximum drawdown of optimized parameter is slightly higher than the traditional parameter for S&P 500. For constituents of Dow Jones, the Sharpe ratio and Sortino ratio remain unchanged, but its maximum drawdown increases 0.01. Therefore, the optimized parameter does not improve the risk return performance, and the maximum loss amount is increased. Compared with the traditional parameter, the signal parameter is longer. Hence, the possible explanation for the maximum drawdown increasing is the result of longer signal parameter. It causes a longer holding period in general. In the stock market, a longer holding period causes risk increase. Based on the comparison result, the optimized parameter for the MACD improves the trading performance in terms of win rate, Sharpe ratio and Sortino ratio in constituents of Nasdaq and constituents of S&P 500. But the trading performance is slightly worse than traditional parameter for constituents of Dow Jones. Meanwhile, the maximum drawdown increases over this dataset. The higher P&L ratio is the result of outlier. Therefore, the improvement of optimized parameter does not come up to my expectations.

Table 3: The comparison result for the MACD$_{crossoversigabout0}$ strategy by adopting traditional parameter and optimized parameter.

|  | NT | WR | P&L | SR | Sortino Ratio | MDD |
|---|---|---|---|---|---|---|
| Panel A: Dow Jones | | | | | | |
| (12,26,9) | 217 | 0.50 | 2.92 | 0.34 | 0.50 | 0.36 |
| (9,27,40) | 217 | 0.49 | 3.65 | 0.34 | 0.50 | 0.37 |
| Panel B: Nasdaq | | | | | | |
| (12,26,9) | 697 | 0.52 | 4.74 | 0.50 | 0.75 | 0.43 |
| (18,35,39) | 331 | 0.59 | 8.44 | 0.52 | 0.78 | 0.42 |
| Panel C: S&P 500 | | | | | | |
| (12,26,9) | 3616 | 0.49 | 3.23 | 0.35 | 0.52 | 0.43 |
| (17,31,23) | 2159 | 0.54 | 4.08 | 0.39 | 0.57 | 0.44 |

Figure 6: The scatter plot for the MACD$_{crossoversigabout0}$ strategy by adopting optimized parameter.

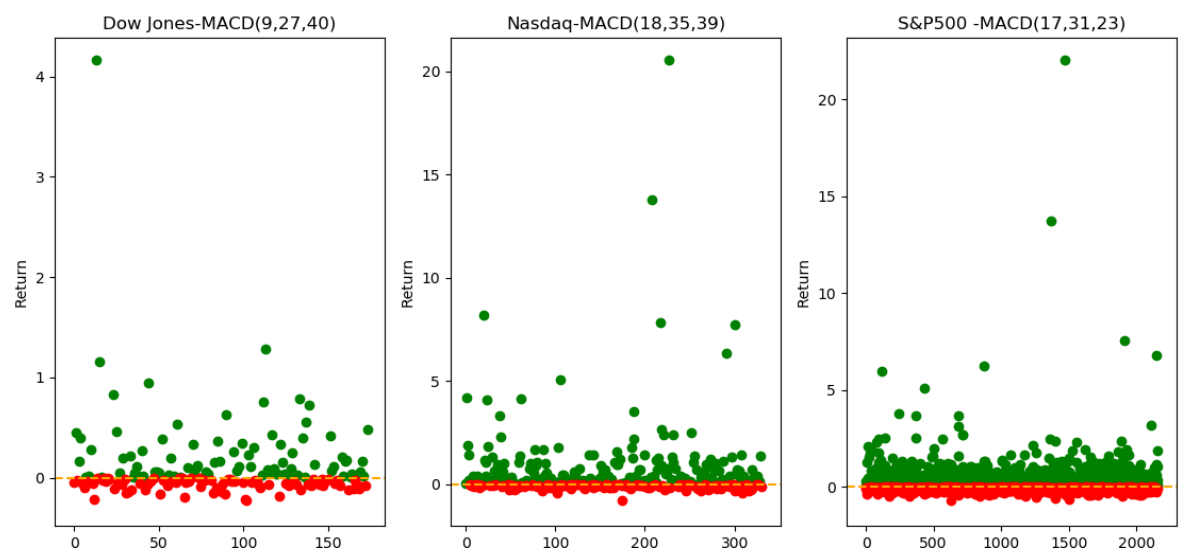

### 3.2.3 Comparison result for MACD strategies with other indicators

Parameters and trading signal identification rule of all combination strategies are specified in Table 4. Based on these combinations, I run the backtesting again to evaluate the trading performance whether the MACD strategy under the traditional parameter (12, 26,9) can be improved by combining with other technical indicators.

Table 5 reports the comparison result. The MACD strategy with RSI (MACD&RSI) and the MACD strategy with MFI (MACD&MFI) perform an overwhelming win rate than others for all data sets. The win rates of MACD&RSI strategy are 0.84, 0.86 and 0.78 respectively, and the win rates of MACD& MFI strategy are 0.69, 0.70 and 0.67 respectively. Additionally, the Sharpe ratio, Sortino ratio and maximum drawdown perform better than other strategies too. Compared with the MACD&MFI strategy, the Sharpe ratio of MACD&RSI strategy performs better. It means that the MACD&RSI strategy has a better risk adjustment performance Therefore, this strategy is preferable. Although the MACD&RSI strategy performs better than the MACD&MFI strategy, the MACD&MFI strategy performs better in terms of P&L. Therefore, the MACD&MFI strategy has a better opportunity of generating a high return on each transaction. That indicates the win rate enhancement of MACD&RSI might be the result of sacrificing higher return opportunities. This explanation corresponds to the Sharpe ratio comparison result because it reveals that investors seek a higher return by taking more risk.

As for MACD&Bollinger band strategy, it improves the win rate like the MACD&MFI strategy and the MACD&RSI strategy did, but the performances are not outstanding. It only performs slightly better than the MACD strategy. Furthermore, the Sharpe ratio and the Sortino ratio are extremely low for Dow Jones and S&P 500 data sets. The risk adjustment performance is poor so that this strategy should not be applied to realistic trades. Lastly, the MACD&SAR strategy has the highest P&L ratio among those strategies for all data sets. The Sharpe ratio performance is almost equivalent to the MACD&MFI and close to MACD& RSI. However, the maximum drawdown performs slightly worse. All in all, MACD&SAR, MACD&MFI and MACD&RSI improve the risk adjustment performance. Interestingly, the best two strategies take the price pressure factor into account. It turns out that the MACD indicator performance can be improved by combining with price pressure factors. Indeed, the MACD formula does not take this factor into consideration. The MACD indicator can only reveal the trending, based on its mathematical definition. As a result of that, this finding seems to be valuable for investors to further investigate the true usage of MACD indicator and its shortcomings.

Table 4: Parameters and trading signal identification rule for combination strategies

| Indicator | Parameter | Trading signal identification rule |
|---|---|---|
| **MACD** | $window_{fast} = 12$<br>$window_{slow} = 26$<br>$window_{sign} = 9$ | Buy: $Macd_t > Signal_t$ & $SMA(BBW, 10)_t > SMA(BBW, 50)_t$<br><br>Sell: $Macd_t < Signal_t$ & $SMA(BBW, 10)_t < SMA(BBW, 50)_t$ |
| **Bollinger Bands** | $window = 14$<br>$window_{stdev} = 2$<br>$window_{ssma} = 10$<br>$window_{lsma} = 50$ | |
| **MACD** | $window_{fast} = 12$<br>$window_{slow} = 26$<br>$window_{sign} = 9$ | Buy: $Macd_t > Signal_t$ & $Sar_{t-1} > Close\ Price_{t-1}$ & $Sar_t < Close\ Price_t$<br><br>Sell: $Macd_t < Signal_t$ & $Sar_{t-1} < Close\ Price_{t-1}$ & $Sar_t > Close\ Price_t$ |
| **SAR** | $\alpha = 0.02$ | |
| **MACD** | $window_{fast} = 12$<br>$window_{slow} = 26$<br>$window_{sign} = 9$ | Buy: $Macd_t > Signal_t$ & $(\forall \{MFI_t, MFI_{t-1}, MFI_{t-2}, MFI_{t-3}, MFI_{t-4}, MFI_{t-5}\} \leq Lower\ Threshold$<br><br>Sell: $Macd_t < Signal_t$ & $(\forall \{MFI_t, MFI_{t-1}, MFI_{t-2}, MFI_{t-3}, MFI_{t-4}, MFI_{t-5}\} \geq Upper\ Threshold$ |
| **MFI** | $window = 14$<br>$MFI_{lower} = 25$<br>$MFI_{upper} = 70$ | |
| **MACD** | $window_{fast} = 12$<br>$window_{slow} = 26$<br>$window_{sign} = 9$ | Buy: $Macd_t > Signal_t$ & $(\forall \{RSI_t, RSI_{t-1}, RSI_{t-2}, RSI_{t-3}, RSI_{t-4}, RSI_{t-5}\} \leq Lower\ Threshold$<br><br>Sell: $Macd_t < Signal_t$ & $(\forall \{RSI_t, RSI_{t-1}, RSI_{t-2}, RSI_{t-3}, RSI_{t-4}, RSI_{t-5}\} \geq Upper\ Threshold$ |
| **RSI** | $window = 14$<br>$RSI_{lower} = 35$<br>$RSI_{upper} = 70$ | |

Table 5: The comparison result of the MACD&Bollinger band, MACD&SAR, MACD&MFI and MACD&RSI.

|  | NT | Win rate | P&L | SR | Sortino ratio | MDD |
| --- | --- | --- | --- | --- | --- | --- |
| Panel A: Dow Jones | | | | | | |
| MACD&BB | 905 | 0.53 | 1.21 | 0.25 | 0.38 | 0.37 |
| MACD&SAR | 653 | 0.51 | 1.91 | 0.39 | 0.59 | 0.27 |
| MACD&MFI | 160 | 0.69 | 1.69 | 0.39 | 0.67 | 0.20 |
| MACD&RSI | 106 | 0.84 | 1.11 | 0.39 | 0.59 | 0.25 |
| Panel B: Nasdaq | | | | | | |
| MACD&BB | 3037 | 0.53 | 1.48 | 0.38 | 0.57 | 0.40 |
| MACD&SAR | 2299 | 0.51 | 2.00 | 0.44 | 0.68 | 0.35 |
| MACD&MFI | 604 | 0.70 | 1.58 | 0.44 | 0.71 | 0.26 |
| MACD&RSI | 334 | 0.86 | 1.40 | 0.50 | 0.83 | 0.27 |
| Panel C: S&P 500 | | | | | | |
| MACD&BB | 15422 | 0.51 | 1.31 | 0.25 | 0.38 | 0.43 |
| MACD&SAR | 11823 | 0.49 | 1.84 | 0.37 | 0.55 | 0.35 |
| MACD&MFI | 3131 | 0.67 | 1.33 | 0.37 | 0.59 | 0.29 |
| MACD&RSI | 1811 | 0.78 | 1.24 | 0.41 | 0.65 | 0.32 |

.

## 3.3. Descriptive Analysis of Strategies' Return

The descriptive statistics of aforementioned strategies in section 3.2 except the MACD strategy that adopts optimized parameter are shown in this section. Reasons for excluding optimized strategy from the descriptive analysis are that there is no single optimized solution for the parameter of MACD and the biased result caused by the longer signal period parameter. The US stock market is the bullish market during the testing period. Therefore, a longer holding leads the return of each trade bias toward positive and produces some outliers as Figure 6 illustrated, due to the bullish market effect.

Table 6 reports the descriptive statistics for aforementioned strategies. The MACD&RSI strategy has the highest mean return for all data sets, and it exhibits positive skewness.

Figure 7 illustrates the return distribution of MACD&RSI strategy. The red line on the picture represents 0. As you observed from the picture, lots of the returns are distributed on the positive side. Thus, the mean of return for each index is greater than zero. In addition, all of them exhibit the positive skewness. This illustration is corresponding to Table 6. Both reveal that the MACD&RSI strategy is profitable, and it is the best strategy among the aforementioned strategies.

Figure 7: The return distribution of the MACD&RSI strategy

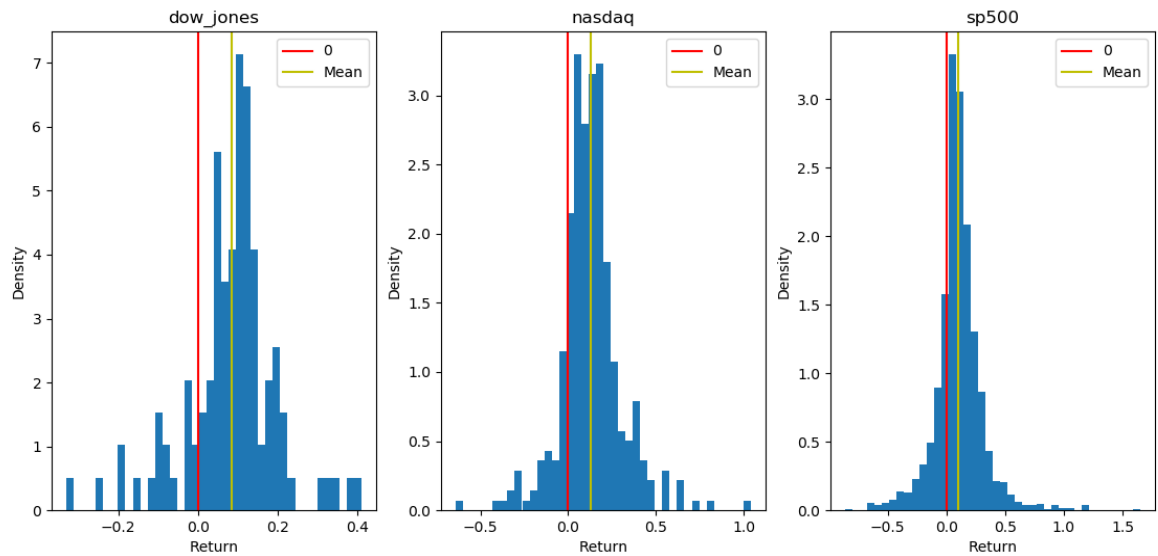

Note: Figure 7 displays the return distribution of the MACD&RSI for the US market indices. The kurtosis value observed from this figure is low. It indicates that there is no outlier that affects the result. Different to the optimized parameters. This strategy is more reliable.

Table 6: Descriptive statistics of aforementioned strategies in different US stock market indices.

|  | Mean Return | Skewness | Kurtosis |
| --- | --- | --- | --- |
| Panel A: Dow Jones | | | |
| MACD&RSI | 0.0839 | -0.4877 | 2.0149 |
| MACD&MFI | 0.0456 | 0.8996 | 3.7057 |
| MACD&SAR | 0.0177 | 1.1829 | 4.5700 |
| MACD&BB | 0.0078 | 0.1066 | 5.9542 |
| $MACD_{crossoversigabout0}$ | 0.0700 | 3.9083 | 23.7921 |
| $MACD_{hist}$ | 0.0076 | -1.8314 | 13.0820 |
| $MACD_{crossoverzero}$ | 0.0120 | 5.3799 | 55.7026 |
| $MACD_{crossoversig}$ | 0.0054 | 1.6862 | 5.5802 |
| Panel B: Nasdaq | | | |
| MACD&RSI | 0.1310 | 0.4755 | 3.9117 |
| MACD&MFI | 0.0610 | 0.8902 | 6.8142 |
| MACD&SAR | 0.0260 | 2.7933 | 21.2636 |
| MACD&BB | 0.0170 | 1.9304 | 19.3021 |
| $MACD_{crossoversigabout0}$ | 0.1750 | 5.9666 | 54.6049 |
| $MACD_{hist}$ | 0.0150 | -0.5426 | 5.6689 |
| $MACD_{crossoverzero}$ | 0.0340 | 5.5020 | 52.2917 |
| $MACD_{crossoversig}$ | 0.0100 | 2.5314 | 19.2918 |
| Panel C: S&P 500 | | | |
| MACD&RSI | 0.1020 | 0.6639 | 6.2549 |
| MACD&MFI | 0.0490 | 0.7654 | 7.5956 |
| MACD&SAR | 0.0180 | 1.8482 | 12.7772 |
| MACD&BB | 0.0100 | 0.7188 | 12.7816 |
| $MACD_{crossoversigabout0}$ | 0.0900 | 6.3119 | 80.4183 |
| $MACD_{hist}$ | 0.0090 | -1.8530 | 12.7160 |
| $MACD_{crossoverzero}$ | 0.0180 | 6.1113 | 80.2034 |
| $MACD_{crossoversig}$ | 0.0080 | 2.9540 | 24.6263 |

## 3.4. Accumulated profit measurement

Accumulated profit measures the arbitrage ability of the strategy. Usually, the price movement of stocks is fluctuating. The price will not always go up or go down. Therefore, if the strategy can seize the opportunity of price movement to make the right decision, it means that selling the stock before it drops and buying back before it rises, the accumulated profit should be high.

Table 7: Accumulated profit measurement

|  | Gain | Loss | AP |
|---|---|---|---|
| Panel A: Dow Jones | | | |
| $MACD_{crossoversigabout0}$ | 2071660.14 | -715515.58 | 1356144.56 |
| MACD&SAR | 2407224.15 | -1192173.84 | 1215050.31 |
| $MACD_{hist}$ | 3178258.88 | -2168256.92 | 1010001.96 |
| $MACD_{crossoversig}$ | 3731022.65 | -2721919.39 | 1009103.26 |
| $MACD_{crossoverzero}$ | 2586154.52 | -1718223.69 | 867930.83 |
| MACD&RSI | 976732.37 | -167825.71 | 808906.66 |
| MACD&MFI | 889412.16 | -239047.00 | 650365.16 |
| MACD&BB | 2270547.15 | -1645944.38 | 624602.77 |
| Panel B: Nasdaq | | | |
| $MACD_{crossoversigabout0}$ | 16537133.38 | -3229504.91 | 13307628.47 |
| $MACD_{crossoverzero}$ | 20476642.67 | -8634751.57 | 11841891.10 |
| $MACD_{hist}$ | 19188702.00 | -11162249.39 | 8026452.61 |
| $MACD_{crossoversig}$ | 20387706.93 | -13436706.00 | 6951000.93 |
| MACD&SAR | 13054336.44 | -6173853.52 | 6880482.92 |
| MACD&Bollinger Bands | 12882828.73 | -7633448.76 | 5249379.97 |
| MACD&RSI | 4523055.82 | -540568.95 | 3982486.87 |
| MACD&MFI | 4926214.28 | -1369457.89 | 3556756.39 |
| Panel C: S&P 500 | | | |
| $MACD_{crossoversigabout0}$ | 46777630.24 | -15267366.75 | 31510263.49 |
| $MACD_{crossoverzero}$ | 59271832.62 | -33425260.05 | 25846572.57 |
| $MACD_{crossoversig}$ | 82302654.30 | -56606660.74 | 25695993.56 |
| MACD&SAR | 47548043.27 | -26811264.42 | 20736778.85 |
| $MACD_{hist}$ | 68020061.19 | -49190919.65 | 18829141.54 |
| MACD&RSI | 21696310.58 | -4972917.39 | 16723393.19 |
| MACD&MFI | 21827355.23 | -7910847.07 | 13916508.16 |
| MACD&BB | 48261406.92 | -35432446.81 | 12828960.11 |

Table 7 reports the accumulated profit result of different MACD-based strategies in data sets. The result is quite interesting. Although the MACD&RSI strategy has a good win rate, better average return and risk adjustment return, the accumulated profit is ranked in the last third. In contrast, the MACD$_{crossoversigabout0}$ strategy is ranked in the top. It indicates that the MACD&RSI strategy is a good strategy, but it is not good enough to seize the arbitrage opportunity. Indeed, it misses some arbitrage opportunities so that the accumulated profit is low. Because the shorting behavior is not considered in this backtesting program, the possible mistakes of missing arbitrage opportunities that are selling the stock early or buy the stock late during the uptrend.

## 3.5. Summary

In this section, I evaluate the performance of different MACD-base strategies through a backtesting system, which is programmed in Python. Based on the Table 5 report, the MACD&Bollinger band strategy is the most undesirable strategy from all aspects. It has the lowest Sharpe ratio and Sortino ratio. The MACD&RSI strategy outperforms others from the win rate, average return and risk adjustment return aspects. However, this strategy leads investors to miss some arbitrage opportunities according to the Table 7 report. The MACD&MFI strategy is the second preferred strategy in terms of the same aspect. The win rate of this strategy is over 0.6 for all data sets. However, it has the same characteristic of the MACD&RSI strategy. It misses some arbitrage opportunities too. The MACD$_{crossoversigabout0}$ gives the highest accumulated profit to investors, but the risk adjustment return performance is unfavorable. Investors bear higher risk to earn slightly better returns.

As for the MACD strategy that adopts optimized parameter, Figure 8 illustrates that the return is outstanding. It is better than other strategies. However, from the result of GA, there is no single optimized parameter for the MACD applied to all the US stock market indices. The optimized parameter for each US stock market index is different. Notice that the common things in the optimized parameter of different US stock market indices are long signal parameters. The US stock market is almost in a bullish state from 2015 to 2021. A longer signal parameter causes investors to hold the stock position for a longer period. As a result of that, this situation leads the return bias to positive. Based on the Figure 6, the optimized parameter generates some outliers of return which weakens the accuracy of prediction. Therefore, the solution that is solved by the GA is the local optimal solution for the specific US stock market index for the specific period only. It probably is the overfitting result. Therefore, I do not advocate using the optimized parameter to realistic trade.

Interestingly, only the MACD-strategies that take the price pressure indicator like RSI and MFI into consideration can produce an outstanding win rate and maximum drawdown based on the Table 5 report. The RSI and the MFI seem to cover the MACD shortcoming which ignores the price pressure information. Although these two indicators generate small accumulated profit, the price pressure factor is a valuable finding to investors. In addition, the price pressure can be explained by trading volume in the market. As a result of that, I use the trading volume as a crucial element for improving the MACD indicators in section 4.

Figure 8: The return distribution of $MACD_{crossoversigabout0}$ by adopting optimized parameter

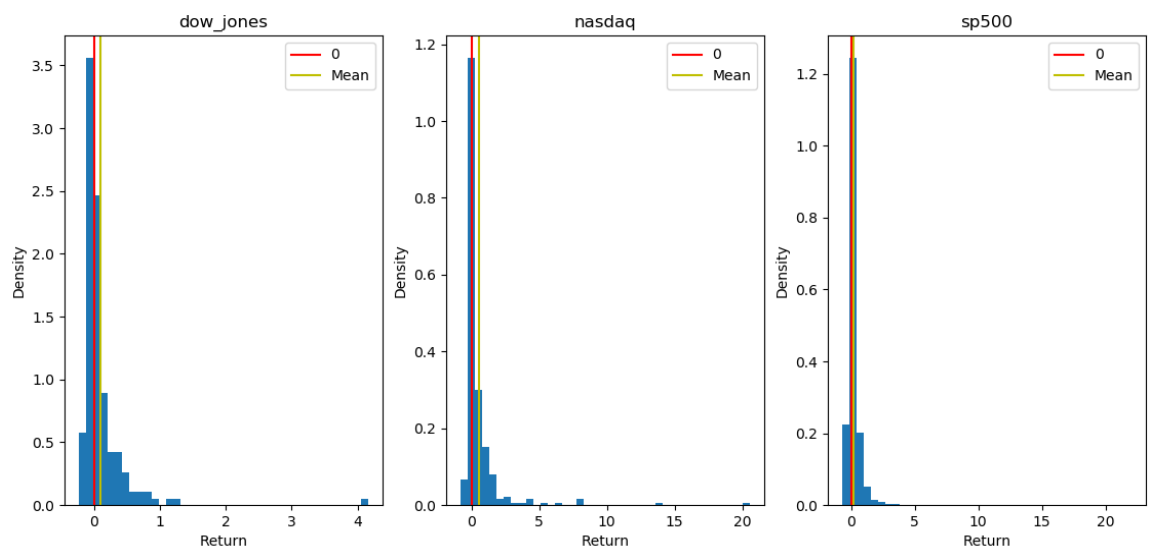

## 4. Comprehensive Technical Indicators

The previous result shows that considering the price pressure factor can improve the win rate and the risk adjustment return performance for the MACD strategy. Therefore, there is a possibility for improving the MACD indicator in a mathematical way instead of combining it with other indicators to build a trading strategy simply. Because the combination method causes the accumulated profit decreased. It is evidenced by results of the MACD&RSI strategy and MACD&MFI strategy in the previous section. Therefore, I derived a new indicator that can improve the comprehensive trading performance, by adding the volume or other factors that are related to volume. The new indicator is called the volume price volume moving average (VPVMA). The VPVMA is more sensitive to the current price change than the MACD. The sensitivity is very important to the technical indicator because it represents the reflection ability. Usually, a good technical indicator should be sensitive to the price change in order to capture the market information accurately. That's why I combine volume and price together to improve the MACD formula. Based on the improvement of sensitivity, the accumulated profit decreasing problem, which existed after the price pressure indicator was combined with the MACD strategy, has been solved as well.

## 4.1. Volume Price Volume Moving Average (VPVMA)

This indicator is the derivation of MACD because it takes the volume, the typical price and the daily volatility into consideration, but does not completely change the fundamental mathematical steps dramatically. The indicator includes two moving average series like the MACD. The first series is the VPVMA, and the second is the VPVMAS. The VPVPMS is the simple moving average of VPVMA, but the computation of VPVMA series is slightly more complicated than the MACD indicator. It uses the typical price instead of the close price to calculate the volume weighted price. Typical price is the arithmetic average of the high price, low price and closing price of a stock. Because stock has different market prices in each trading session every day, using the arithmetic average of price can consider all the price information in a trading decision. After that, I apply the volume weighted method to the typical price in order to calculate the short-term and long-term volume weighted price moving average respectively. These two moving average series are calculated by the typical price, which takes the comprehensive price information into account. Therefore, the volume weighted price moving average should be the most accurate series to capture the price moment in the given period. Beside the typical price, the daily price volatility is needed. To study the daily price volatility, which is the standard deviation of stock price in one day, can help the investor to capture the short-term volatility and identify the price breakthrough. Furthermore, I apply the exponential moving average to smooth the product of volume weighted price moving average and the daily price volatility so that the final result of VPVMA is smooth enough, and false signals are reduced. The following are mathematical steps and formulas of the VPVMA indicator.

$$Typical\ Price(TP) = \frac{High + Low + Close}{3} \qquad (4.2\text{-}1)$$

$$SVWMA = \frac{\sum_i^n TP_i * V_i}{\sum_i^n V_i} \ where\ n = short\ term\ period \qquad (4.2\text{-}2)$$

$$LVWMA = \frac{\sum_i^n TP_i * V_i}{\sum_i^n V_i} where\ n = long\ term\ period \qquad (4.2\text{-}3)$$

$$Daily\ Volatility(DV) = Std(High, Low, Close, Open) \qquad (4.2\text{-}4)$$

$$\text{ESVMap} = \text{EMA}(\text{SVWMA} * \text{DV, s})\ where\ s = short\ term\ period \qquad (4.2\text{-}5)$$

$$\text{ElVMap} = \text{EMA}(\text{SVWMA} * \text{DV, s})\ where\ s = long\ term\ period \qquad (4.2\text{-}6)$$

$$VPVMA = ESVMap - ELVMap \qquad (4.2\text{-}7)$$

$$VPVMAS = \frac{\sum_i^n VPVMA_i}{n}\ where\ n = signal\ period \qquad (4.2\text{-}8)$$

The above formula shows that the VPVMA indicator is more elaborate and complex than the MACD mathematically. It takes price factor, time factor, volume factor and price volatility factor into account so that the sensitive moment of price and time trend can be captured correctly.

## 4.2. Results

I perform the backtesting to evaluate the VPVMA indicator. Parameters and the trading signal identification method is shown on Table 8

Table 8: Parameters and trading signal identification rule for the VPVMA indicator

| Indicator | Parameter | Trading signal identification rule |
|---|---|---|
| **VPVMA** | $Window_{fast} = 12$<br>$Window_{slow} = 26$<br>$window_{sign} = 9$<br>bandwidth= 0.1 | Buy: (VPVMAt > (1+band width) * VPVMASt) & (VPVMAt-1 <=VPVMASt-1)<br><br>Sell: (VPVMAt < (1-band width *2) VPVMASt) & (VPVMAt-1 <= VPVMASt-1) |

Note: I add a bandwidth parameter in the identification rule to further reduce the false signal. The above conditions show that a buy signal is generated when the VPVMA is greater than VPVMS for a certain percentage. A sell signal is generated when the VPVMA is less than VPVMS for a certain percentage. The bandwidth is a control variable, which means that you can adjust the value to achieve your expected number of trades.

Table 9: Comparison results of VPVMA with selected strategies

|  | NT | AP | Win | P&L | SR | Sortino | MDD |
|---|---|---|---|---|---|---|---|
| Panel A: Dow Jones | | | | | | | |
| MACD$_{crossoversig}$ | 1939 | 1009103.26 | 0.41 | 1.97 | 0.33 | 0.51 | 0.28 |
| MACD$_{crossoversigabout0}$ | 217 | 1356144.56 | 0.50 | 2.92 | 0.34 | 0.50 | 0.36 |
| MACD&RSI | 106 | 808906.66 | 0.84 | 1.11 | 0.39 | 0.59 | 0.25 |
| MACD&MFI | 160 | 650365.16 | 0.69 | 1.69 | 0.39 | 0.67 | 0.20 |
| VPVMA | 602 | 1501820.08 | 0.60 | 1.41 | 0.49 | 0.76 | 0.25 |
| Panel B: Nasdaq | | | | | | | |
| MACD$_{crossoversig}$ | 6485 | 6951000.93 | 0.41 | 2.15 | 0.44 | 0.67 | 0.35 |
| MACD$_{crossoversigabout0}$ | 697 | 13307628.47 | 0.52 | 4.74 | 0.50 | 0.75 | 0.43 |
| MACD&RSI | 334 | 3982486.87 | 0.86 | 1.40 | 0.50 | 0.83 | 0.27 |
| MACD&MFI | 604 | 3556756.39 | 0.70 | 1.58 | 0.44 | 0.71 | 0.26 |
| VPVMA | 2266 | 7835408.86 | 0.58 | 1.68 | 0.52 | 0.80 | 0.32 |
| Panel C: S&P 500 | | | | | | | |
| MACD$_{crossoversig}$ | 32894 | 25695993.56 | 0.40 | 2.15 | 0.38 | 0.58 | 0.34 |
| MACD$_{crossoversigabout0}$ | 3616 | 31510263.49 | 0.49 | 3.23 | 0.35 | 0.52 | 0.43 |
| MACD&RSI | 1811 | 16723393.19 | 0.78 | 1.24 | 0.41 | 0.65 | 0.32 |
| MACD&MFI | 3131 | 13916508.16 | 0.67 | 1.33 | 0.37 | 0.59 | 0.29 |
| VPVMA | 12436 | 27899282.75 | 0.56 | 1.55 | 0.45 | 0.69 | 0.33 |

Compared with all MACD$_{crossoversig}$ strategy and MACD$_{crossoversigabout0}$ strategy, the win rate, the Sharpe ratio, the Sortino ratio and Maximum drawdown are all better. In addition, the excessive trading problem existed MACD$_{crossoversig}$ is eliminated successfully, due to the effect of double EMA technique and filter bandwidth applied to the VPVMA. Although the number of trades is larger than MACD$_{crossoversigabout0}$ for all data sets, the better risk adjustment return is more preferable in the stock market. There is no reason to give up an indicator that can produce a better risk adjustment return just because it generates a slightly higher number of trades. Lastly, the accumulated profit is lower than MACD$_{crossoversigabout0}$ for Nasdaq and S&P 500 data sets. However, those returns are the result of investors bearing more risk evidenced by the Sharpe ratio, the Sortino ratio, maximum drawdown values. But their difference is not as large as MACD&RSI strategy and MACD&MFI strategy. I consider it is acceptable. In a comparison with the MACD&MFI strategy and MACD&RSI strategy, the risk adjustment return outperforms others too. Although the win rate of this strategy cannot achieve as great as MACD&RSI and MACD&MFI strategies, the accumulated profit can make up for the slightly lower win rate.

Figure 9: The return distribution of the VPVMA strategy

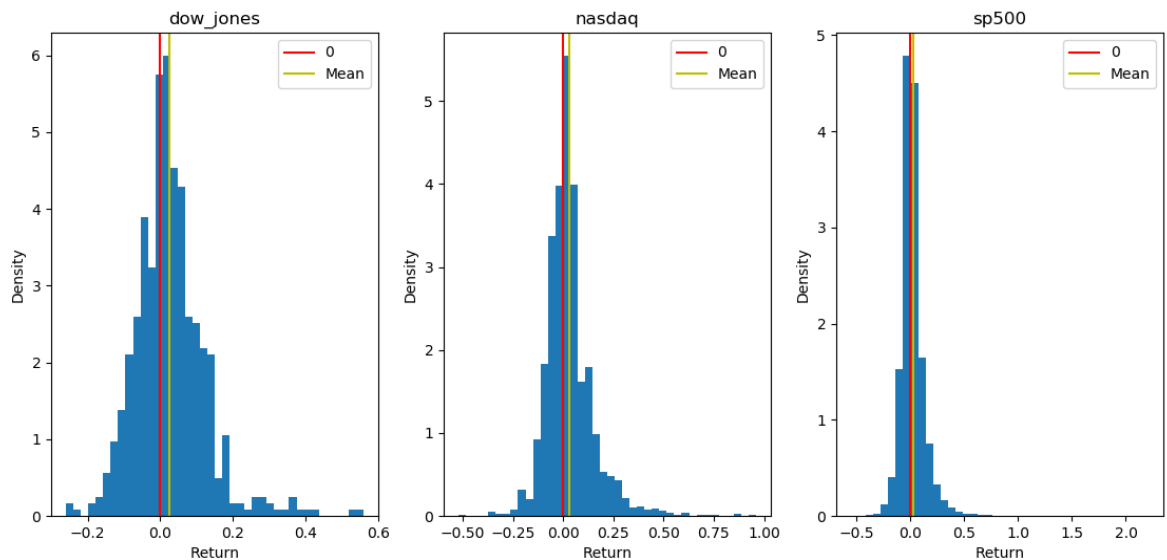

Note: Figure 9 displays that the return of VPVMA of each trade concentrates on the range (-0.2 – 0.2). And the Kurtosis observed from this figure is normal. It indicates that the return of this indicator is stable and predictable.

# 5. Conclusion

In this thesis, I evaluate different MACD-based strategies. The evaluation includes MACD strategies with different signal identification rules, MACD strategies combined with different technical indicators and the VPVMA strategy, which is derived by MACD. In addition, I use the GA algorithm to identify the optimized parameters for MACD. Unfortunately, the GA does not identify the same optimized parameters for the different US stock market indices (i.e.., Dow Jones, Nasdaq and S&P 500). It is an indication that no optimized parameters existed in the MACD. I think that the result is reasonable and valid. According to the definition of the MACD, it is a signal processing technique. The crossover between the MACD line and signal line indicates the signal impulse. It is normal that different parameters cause different signal impulses. Furthermore, the signal impulse originally does not reflect the appropriate time but reflects the volatility information. Any impulse indicates the stock market is fluctuating. Investors only can find the clue from the impulse to justify the trend and identify an arbitrage opportunity. Therefore, trying to identify the optimized parameter through machine learning, genetic algorithm and exhaustion search is unreasonable. Any optimized results generated from the above process is only the best result for the observation period. It is not the global optimal solution. According to the backtesting results from the previous section, some strategies can generate a decent profit. The MACD&MFI strategy and MACD&RSI strategy have a high win rate characteristic, but the profitability is unattractive. These two strategies are ranked in the last three from the profitability aspect. Both cannot capture the arbitrage opportunity as much as possible from the volatility. The VPVMA strategy and the $MACD_{crossoversigabout0}$ strategy can generate attractive profit. But the comparison results show that the VPVMA strategy is a more valuable one. Besides, the win rate is better, the risk adjustment return in terms of the Sharpe ratio, the Sortino ratio and the maximum drawdown outperform others too. Hence, the VPVMA strategy is the most preferable strategy, and it is an indication that the mathematical design of this new indicator is successful.

From the panel data comparison, I found that NASDAQ dataset always perform better than Dow Jones and S&P 500 when investors use the MACD-based strategies. The result is because of their constituents. NASDAQ data set consists of 100 tech-start companies, which are invested heavily by the organization investors, and is traded actively by the indie investors too. It indicates that the liquidity of NASDAQ is better than others. But the Nasdaq constituents are high growth stocks. Therefore, their market capitalization is fluctuated. That leads the index is unstable because NASDAQ index is a market capitalization-weighted index. As a result of that, assets' price might not be reflected timely, and investors can use the market information to find the best trading opportunity. It means that the NASDAQ is inefficient market compared with Dow Jones and S&P 500.

All in all, the MACD&MFI strategy, the MACD&RSI strategy, the MACD$_{crossoversigabout0}$ strategy and the VPVMA strategy can generate a positive return. And the VPVMA is the best among them. Although I have not compared the return of those strategies to other different trading type strategies like buy-hold strategy and technical pattern recognition strategy in this thesis, a positive risk-versus return is an indication that those MACD-based strategies are valuable to investors. Further comparing those strategies with benchmark, other type strategies, or studying the return series characteristic are the future research direction.